\documentclass[prl,twocolumn,superscriptaddress,showpacs,floatfix]{revtex4-1}
\usepackage{graphicx,latexsym,amssymb,bm}
\begin{document}
\newcommand{\mf}{{\bf f}}
\newcommand{\mA}{{\bf A}}
\newcommand{\mB}{{\bf B}}
\newcommand{\mC}{{\bf C}}
\newcommand{\mF}{{\bf F}}
\newcommand{\mE}{{\bf E}}
\newcommand{\mH}{{\bf H}}
\newcommand{\mI}{{\bf I}}
\newcommand{\mJ}{{\bf J}}
\newcommand{\mL}{{\bf L}}
\newcommand{\mM}{{\bf M}}
\newcommand{\mN}{{\bf N}}
\newcommand{\mG}{{\bf G}}
\newcommand{\mP}{{\bf P}}
\newcommand{\mR}{{\bf R}}
\newcommand{\mS}{{\bf S}}
\newcommand{\mU}{{\bf U}}
\newcommand{\mfe}{{\bf f_{\eta}}}
\newcommand{\mh}{{\bf h}}
\newcommand{\mg}{{\bf g}}
\newcommand{\mgi}{{\bf g^{-1}}}
\newcommand{\mk}{{\bf k}}
\newcommand{\md}{{\bf d}}
\newcommand{\mm}{{\bf m}}
\newcommand{\mn}{{\bf n}}
\newcommand{\mv}{{\bf v}}
\newcommand{\mq}{{\bf q}}
\newcommand{\mr}{{\bf r}}
\newcommand{\mun}{{\bf u}}
\newcommand{\munit}{{\bf 1}}
\newcommand{\mzero}{{\bf 0}}
\newcommand{\msigma}{\boldsymbol{\sigma}}
\newcommand{\mkappa}{\boldsymbol{\kappa}}
\newcommand{\mpi}{\boldsymbol{\pi}}
\newcommand{\mepsilon}{\boldsymbol{\epsilon}}
\newcommand{\mOmega}{\boldsymbol{\Omega}}
\newcommand{\mroi}{{\bf r_{01}}}
\newcommand{\mrio}{{\bf r_{10}}}
\newcommand{\mrto}{{\bf r_{20}}}
%
%
\newcommand{\bra}[1]{\langle{#1}|}
\newcommand{\brar}[1]{\langle{#1}\|}
\newcommand{\ket}[1]{|{#1}\rangle}
\newcommand{\ketr}[1]{\|{#1}\rangle}
\newcommand{\dbar}[1]{\overline{\overline{{#1}}}}
%
%
\newcommand{\braket}[2]{\langle {#1}|{#2}\rangle}
\newcommand{\ms}[2]{\normalsize{#1}\small{#2}\normalsize}
\newcommand{\col}[2]{\left(\begin{array}{c}{#1}\\[1mm]{#2}
                     \end{array}\right)}
%
%
\newcommand{\mat}[4]{\left(\begin{array}{cc}{#1}&{#2}\\[1mm]{#3}&{#4}
                     \end{array}\right)}
\newcommand{\matn}[4]{\left(\!\begin{array}{cc}{#1}\!&{#2}\\[1mm]{#3}\!&{#4}
                     \end{array}\!\right)}
\newcommand{\matnn}[4]{\left(\!\!
                      \begin{array}{cc}{#1}\!\!&{#2}\\[1mm]
                                       {#3}\!\!&{#4}
                      \end{array}\!\!\right)}
\newcommand{\mao}[4]{\begin{array}{cc}{#1}&{#2}\\[1mm]{#3}&{#4}
                     \end{array}}
\title{Photon Polarization Precession Spectroscopy for
  High-Resolution Studies of Spinwaves}
\author{Ralf R\"ohlsberger}
\affiliation{Deutsches Elektronen-Synchrotron DESY, 
Notkestr. 85, 22607 Hamburg, Germany}
%
\date{\today}
\begin{abstract}
A new type of spectroscopy for high-resolution studies of
spin waves that relies on resonant scattering of hard x-rays is introduced.
The energy transfer in the scattering process is
encoded in the precession of the polarization vector of the scattered photons. Thus, the energy resolution of
such a spectroscopy is independent of the bandwidth of the probing
radiation. The measured quantity resembles the intermediate scattering
function of the magnetic excitations in the sample.
At pulsed x-ray sources, especially x-ray lasers, the proposed
technique allows to take single-shot spectra of the magnetic
dynamics. The method opens new avenues to study low-energy
non-equilibrium magnetic processes in a pump-probe setup.
\end{abstract}
\pacs{07.85.Nc, 78.70.Ck, 75.30.Ds, 75.25.-j}
\maketitle
\begin{figure}[b]
\begin{center}
\includegraphics[width=0.5\textwidth]{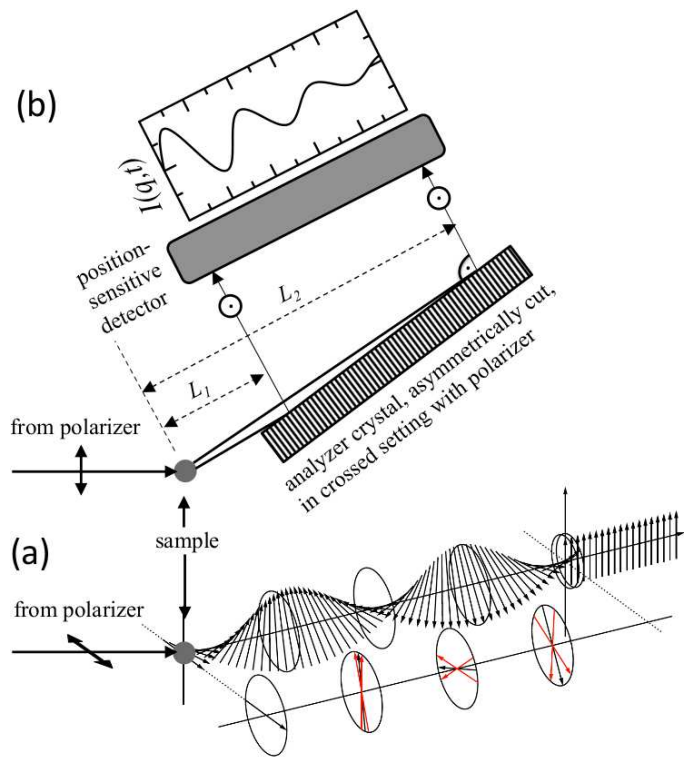}
\caption[]{
(a) The photon polarization of the x-rays scattered with momentum
transfer $q$ from the spinwave
excitation (see Fig. 1) performs a precessional
motion in space with a spatial period of $S = \pi c/\Omega$.
The scattered photons are analyzed with a vertical polarizer that projects the
vertical component of the scattered radiation, revealing the
intermediate scattering function $I(q,t=L/c)$ as function of the distance $L$
from the sample. (b) Employing an asymmetrically cut analyzer crystal
(Bragg angle $\Theta_B \approx 45^{\circ}$) the function can be
recorded within a time interval ($L_1/c, L_2/c$) on a position
sensitive detector, potentially within a single shot of an intense
pulsed x-ray source. A distribution of the spinwave frequencies in the
sample leads to a fanning out of the photon polarization vectors (red arrows) and
thus a decreasing modulation amplitude with increasing distance from
the sample.}
\label{polspiral1}
\end{center}
\end{figure}

The enormous potential of fast spin manipulation for applications in information storage,
processing and retrieval stimulates a growing interest in the excited
states and non-equilibrium properties of magnetic structures. 
The elementary quanta of excitations in an ordered ensemble of magnetic moments are magnons, also
known as spinwaves when described classically.
Magnetic excitations are of particular interest in magnetic
systems with competing interactions. For example, geometrically frustrated magnets
exhibit persistent magnetic excitations even at lowest temperatures
with most of their spectral weight shifted towards low energies
\cite{GDG*99,Ehl06}. This has been shown for crystalline systems \cite{DAA*11,BSS12} and
remains an interesting subject to be studied in artificially
structured systems \cite{MSL*10,MHR*11}. 
Moreover, the emerging field of spinwave engineering a.k.a magnonics \cite{NG09,KDG10,LUGM11,DS13}
relies on the understanding of low-energy magnetic excitations in nanostructured
systems.\\
For a precise measurement of the magnetic excitation spectrum,
high-resolution spectroscopic techniques are required.
In the optical regime, Brillouin light scattering allows to probe
magnons with outstanding energy resolution \cite{SW79,HBG89,Hil00}. 
The  use of visible light, however, prevents the
access to high momentum transfers and thus limits the range of
accessible length scales. Dimensions down to interatomic distances can
be reached via resonant inelastic x-ray scattering (RIXS) or
inelastic neutron scattering (INS). While single-magnon
spectroscopy with x-rays has been demonstrated just
recently \cite{BBB*10}, inelastic neutron scattering for magnon studies is an
established technique since decades.
In all these methods the energy resolution is determined (and limited) by
the energy spread of the incoming particles (assuming a perfect analyzer), thus further bandwidth
reduction to achieve better energy resolution goes at the expense of
the signal to noise ratio. Due to limited instrumental resolution RIXS
is restricted to energy transfers above $\approx$ 50 meV, so that the
low-energy regime of magnetic excitations is still the domain of INS. \\
In fact, a very elegant decoupling of the energy
resolution from the bandwidth of the probing particles has been
achieved in the method of neutron spin echo (NSE) spectroscopy
\cite{Mez72}. 
Small energy transfers upon inelastic scattering are encoded as phase
shift in the precessing polarization of the neutrons. In combination with momentum-resolved
triple-axis spectrometry \cite{Pyn78}, the dynamical structure factor
of magnons can be effectively probed with $\mu$eV resolution \cite{BKHK06,NKM*11} although the
energy bandwidth of the incident neutrons is much larger. As a time-of-flight
method, the neutron spin echo technique relies on the finite rest mass
of the neutron. Therefore, at first sight, this technique does not seem to be
directly transferable to photons.\\
%
%
\begin{figure}[t]
\begin{center}
\setlength{\unitlength}{1cm}
\includegraphics[width=0.5\textwidth]{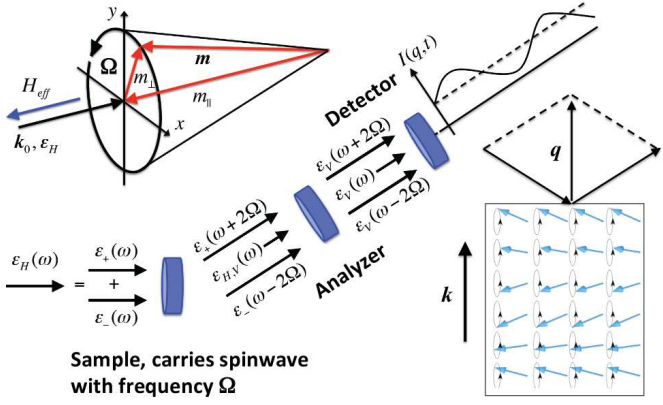}
\caption[]{
Resonant scattering of linearly polarized x-rays from a sample that carries a spinwave represented by
a precessional motion with frequency $\Omega$ shown in the upper left. X-ray magnetic linear dichroism
lets the spinwave act as a rotating half-wave plate that imposes
frequency shifts of +/- 2$\Omega$ on the right/left-circular
polarization components of the scattered x-rays, respectively. After
passage through a linear analyzer one observes a signal that is
modulated with frequency 2$\Omega$. In the general case of an
arbitrary spinwave spectrum the signal is
proportional to the intermediate scattering function $S(q,t)$ that
probes magnetic dynamics over correlations distances $\lambda_S =
2\pi/q$. }
\label{polprecess}
\end{center}
\end{figure}
In this Letter we introduce a new type of inelastic x-ray spectroscopy
to probe magnetic excitations that exhibits
similarities to NSE, the basic principle of which is illustrated in Fig.\,\ref{polprecess}. 
The technique described here relies on resonant magnetic scattering of x-rays
in the presence of x-ray linear dichroism (XMLD). Under these
conditions a magnetic sample
with a spinwave of frequency $\Omega$ exhibits exhibits properties of a half-wave
plate rotating with frequency $\Omega$. A half-wave plate
reverts the \textit{angular} momentum of incident circularly
polarized light, thus it constitutes the angular variant of a mirror that reverses the
\textit{linear} momentum of the light that is backreflected from it. In the
same fashion as photons reflected from a moving mirror experience a
linear Doppler shift of $\Delta E = \mq\cdot\mv$, the photons transmitted
through a rotating half-wave plate experience an angular Doppler
effect \cite{SW78,GA79,Gar81,BGS*94,BB97,CRD*98} of $\Delta E =
\mL\cdot\mOmega$, with $\mq$ and $\mL$ being linear and
angular momentum transfer, and $\mv$ and $\mOmega$ being linear and
angular velocity, respectively. Note, that a necessary
condition for the linear and angular Doppler effect is either a broken
translational invariance (interfaces, lattice planes) or a broken
rotational invariance (anisotropy, optical or magnetic axes),
respectively. While the linear Doppler effect forms the basis for
vibrational spectroscopies, we will exploit here the angular Doppler
effect for a new type of spectroscopy in the x-ray regime to probe
spinwaves within a frequency range reaching up to 100 Ghz.\\
In the following we evaluate the scattering of linearly polarized
x-rays from magnetic samples that carry spinwave excitations.  
We first concentrate on small momentum transfers $\mq$ so that we can use the
forward scattering amplitude to describe the scattering process that
will be treated in the kinematical approximation. The spinwave is represented by a
magnetization vector $\mm = \mm(t)$ that precesses on a cone around the
direction of the effective magnetic field, as illustrated in
Fig.\,\ref{polprecess}. 
The sample and optical elements like polarizers are
described in the Jones matrix formalism
by (2$\times$2) matrices for a given polarization basis (e.g., linear or circular) with unit vectors
$(\mepsilon_{\mu},\mepsilon_{\nu})$. 
We assume the photon energy being tuned to a resonance
(electronic or nuclear) that is sensitive to the sample magnetization
which lifts the degeneracy of the magnetic sublevels due to a
spin-orbit interaction of electronic levels or a magnetic hyperfine
interaction of nuclear levels. \\
In an resonant inelastic scattering experiment at photon energy 
$E = \hbar\omega_0$, the intensity scattered into an energy
interval $dE$ and solid angle $d\Omega$ 
is proportional to the double differential cross section\,:
\begin{equation}
\label{intensity}
I_{fi} \sim \frac{d^2\sigma}{dE\,d\Omega} = |f(\mq,\omega,\omega_0,\epsilon_f,\epsilon_i)|^2\,S(\mq,\omega)
\end{equation}
where $f(\mq,\omega,\omega_0,\epsilon_f,\epsilon_i)$ is the coherent
atomic scattering
amplitude for a given energy transfer $\hbar\omega$, momentum transfer 
$\mq = \mk_f - \mk_i$ and change of polarization $\epsilon_i
\rightarrow \epsilon_f$. $S(\mq,\omega)$ is the dynamical structure
factor of the spinwave with frequency $\omega$ and momentum $\mq$. 
$f$ is derived from a $2\times 2$ matrix that accounts for the
polarization dependence of the scattering process\,:
\begin{equation}
\label{secondeq}
\mf(\mq,\omega) = \frac{2\pi}{k_0}\sum_i\,\varrho_i \mM_i(\mq,\omega)\,.
\end{equation}
where the sum runs over all atomic species in the sample. For
simplicity we drop the $\mq$ dependence in the following and assume
that the scattering proceeds close to the forward direction. 
$\mM_i$ is then the 2$\times$2 matrix of the coherent forward scattering length of the
$i^{th}$ atomic species and $\varrho_i$ is the number density of these atoms.
It is convenient to decompose $\mM(\omega)$ into a non-resonant
part $\mE(\omega)$ that describes electronic charge scattering (see
supplementary material \cite{supplement})
and a part $\mN(\omega)$ that contains the contributions from
resonant scattering\,:  
\begin{equation}
\label{scattm}
\mM(\omega) = \mE(\omega) + \mN(\omega)\,.
\end{equation}
With $\mm$ denoting the unit vector of the magnetization at the
position of the atom, the resonant atomic scattering length $\mN(\omega)$ for an electric dipole transition
($L$ = 1) is typically written as\,:
\begin{eqnarray}
\label{ampl}
[\mN(\omega)]_{\mu\nu} & = & \frac{3}{16\pi}\left\{
(\mepsilon_{\mu}\cdot\mepsilon_{\nu})\left[F_{+1} + 
F_{-1}\right]\right.\\
& & \hspace*{7mm} 
- \,i\,(\mepsilon_{\mu}\times\mepsilon_{\nu})\cdot\mm\,
  \left[F_{+1} - F_{-1}\right] 
\nonumber\\[1mm]
& & \hspace*{8mm} \left. 
+ \,(\mepsilon_{\mu}\cdot\mm)(\mepsilon_{\nu}\cdot\mm)\,
  \left[ 2F_0 - F_{+1} - F_{-1} \right]\right\}\,.  
\nonumber
\end{eqnarray}
with $F_0, F_{+1}$ and  $F_{-1}$ being the energy dependent oscillator
strengths for resonant transitions between magnetic
sublevels with $\Delta m = -1, 0, +1$. 
The three terms in Eq.\,(\ref{ampl}) represent different 
polarization dependences.
The first term is not sensitive to the sample magnetization. 
The second term describes circular dichroism (XMCD) because it
depends on the difference between the resonant scattering
amplitudes $F_{+1}$ and $F_{-1}$.
Since its polarization dependence is 
$\mepsilon_{\mu}\times\mepsilon_{\nu} = \mk_0$, it describes orthogonal 
scattering between the states in the polarization basis.
The third term is the important one here. It describes x-ray linear magnetic dichroism
(XMLD, see supplementary material \cite{supplement}) and is responsible for the violation of rotational invariance that gives rise
to the angular Doppler effect on which the method relies.\\
It is convenient to express Eq.\,(\ref{ampl}) in terms of $2\times 2$
matrices within a circular polarization basis.
With $C_{\pm} = (3/16\pi)(F_{+1} \pm F_{-1})$ 
and $D = (3/32\pi)(2 F_0 - F_{+1} - F_{-1})$ we obtain for the
scattering matrix (for derivation see
supplementary material \cite{supplement})\,:
\begin{eqnarray}
\label{matrix}
\mM & = & (E + C_+ + m^2_{\perp} D)\,\mI  + 2\,m_{\parallel} C_-\,\mP_F\\
& & - m^2_{\perp} D\,\mP_{1/2}(\phi) \nonumber\\
\mbox{with} & &
\mP_F = \matn{1}{0}{0}{-1},
\mP_{1/2}(\phi) =
i\matn{0}{e^{-i 2\phi}}{- e^{i 2\phi}}{0}
\nonumber
\end{eqnarray}
where $\mI$ is the unit matrix and
$\mP_{1/2}(\phi)$ is the Jones matrix of a half-wave plate with the
fast axis oriented at an angle $\phi$ relative to the horizontal (see
Fig.\,\ref{polprecess}). \\
In a spinwave, i.e., a collective motion of a large number of
atomic magnetic moments in the sample, the magnetization $\mm(t)$ 
performs precessional motion with angular frequency $\mOmega$ around the effective field $\mH_{eff}$ as
illustrated in Fig.\,\ref{polprecess}. In order to describe the
scattering process from such an ensemble of spins, we need to perform
a transformation of the scattering matrix from the co-rotating frame into the fixed laboratory
frame. This is accomplished via the transformation
\begin{equation}
\mM(\mOmega) = \mR(\mOmega)\,\mM(0)\,\mR^{-1}(\mOmega)
\end{equation}
$\mR(\mOmega)$ is the expectation value of the operator that generates the
precessional motion of the magnetization, given by (see supplementary
material \cite{supplement})
\begin{equation}
\mR(\mOmega) = 
\matn{e^{i \mOmega \cdot \mkappa\,t}}
{0}{0}
{e^{-i \mOmega\cdot \mkappa\,t}}
\end{equation}
Applying this transformation to the scattering matrix in
Eq.\,(\ref{matrix}) leaves the first two terms invariant, but changes
and introduces a time dependence of the third one:
\begin{equation}
\mR(\mOmega)\,\mP_{1/2}(\phi)\,\mR^{-1}(\mOmega) =
\mP_{1/2}(\phi_0 + \mOmega\cdot\mkappa\,t),
\end{equation}
i.e. the spinwave acts like a half-wave plate that rotates with
angular velocity $\mOmega\cdot\mkappa$. 
The scalar product accounts for the projection of the spin precession
cone on the incident wavevector. To simplify the following discussion,
we assume that $\mOmega || \mkappa$ 
so that $\mOmega\cdot\mkappa = \Omega$, and set $\phi_0 = 0$.\\
%
%
In experiments with synchrotron radiation the incident field $\mA_0$ is usually
linearly polarized in the horizontal plane, i.e., $\mA_0 = \mA_H =
\mepsilon_H e^{i\omega t}$.
To evaluate the amplitude of the scattered field $\mA_S = \mM\,\mA_H$,
we write the incident horizontal polarization as
superposition of left- and right-circular polarization, i.e., $\mA_H =
(\mA_- + i\mA_+)/\sqrt{2}$ to obtain the following contribution from the third
term in Eq.\,(\ref{matrix})\,:
\begin{equation}
\label{circrep}
\left[\mP_{1/2}(\Omega t)\,\mA_H\right]_{circ} =
i e^{i\,2\Omega\,t}\,\mA_- + e^{- i\,2\Omega\,t} \mA_+.
\end{equation}
The left(right)-circular component of the incident linear polarization was converted
into right(left)-circular polarization and shifted up(down) in
frequency by $2\Omega$. 
This means that the energy transfer $2\Omega$ from the spinwave to the scattered photon is
encoded in the relative frequency shift of the two circular
components. Note that this result is independent of the carrier
frequency $\omega$, so that the effective energy
resolution of the method is decoupled from the frequency bandwidth of
the incident radiation. 
Writing Eq.\,(\ref{circrep}) in the linear basis yields 
\begin{equation}
\label{linrep}
\left[\mP_{1/2}(\Omega t)\,\mA_H \right]_{lin} =
\cos 2\Omega t\,\mA_H - \sin 2\Omega t\,\mA_V
\end{equation}
which shows that the polarization performs a precessional
motion in space, illustrated in Fig.\,\ref{polspiral1}a. 
%
%
Since two orthogonal polarizations do not interfere with each other
this frequency shift is not directly observable. The interference can
be induced, however, if the radiation is analyzed by a linear
polarizer which projects parallel polarization components. 
The scattered field $\mA_S$ behind a vertical analyzer with Jones matrix $\mP_V$ is given by 
$\mA_S = \mP_V \mM \mA_H$. Inserting $\mM$ given by Eq.\,(\ref{matrix})
with $\phi = 2\Omega t$ we obtain (see supplementary material \cite{supplement})\,:
\begin{equation}
\label{vertanalyser}
\mA_S = \left[
i\,m_{\parallel} C_-\,  +
m_{\perp}^2 D\,\sin 2\Omega t
\right] \mA_V =: f(q,\Omega)\,\mA_V
\end{equation}
Thus, the intense, horizontally polarized direct beam that was scattered by the sample
without interaction with the spinwave is completely blocked by the vertical linear
polarizer. The linearly polarized wave that is transmitted by the vertical
analyzer is modulated with a frequency of $2\Omega$. \\
Inserting the above expression for $f(q,\Omega)$ given by
Eq.\,(\ref{vertanalyser}) into Eq.\,(\ref{intensity}) and integrating
over all frequencies of energy loss and energy gain in the scattering
process, we obtain for the intensity observed behind the analyser\,:
\begin{equation}
\label{final_intens}
 I(q,t) 
\approx I_B\,S(q) + I_S\,\int\limits_{0}^{\infty}\left[S(q,\Omega)
  - S(q,-\Omega)\right]\sin 2\Omega t
\,d\Omega
\end{equation}
with $I_B = m^2_{\parallel} |C_-|^2$ and $I_S = 2 m_{\parallel} m_{\perp}^2 \mbox{Re}\left[ i\,C_- D^{\ast}\right]$,
where we have used that $|m_{\perp}^2 D|^2 \ll |m_{\parallel} C_-|^2$ and
that $\int S(q,\Omega) d\Omega = S(q)$ being the static structure
factor for given $q$. 
The first term in Eq.\,(\ref{final_intens}) is
independent of $t$ and thus contributes a constant background to the
measured intensity. The second term contains information about the
spinwave dynamics. It resembles the intermediate scattering function
as it is obtained in neutron spin echo spectroscopy \cite{Mez72}.
This quantity can also be obtained via time-domain
interferometry based on nuclear resonant scattering \cite{BFM*97} but the
resolution and signal strength of this method
is governed by the intrinsic bandwidth of the nuclear resonant
scattering method.\\
For symmetric functions $S(q,\Omega)$, however, the
second term in Eq.\,(\ref{final_intens}) vanishes. This is typically the
case in the limit $\hbar\Omega \ll k_B T$ under the condition of
detailed balance at thermal equilibrium, i.e., $S(q,-\Omega) =
\exp(-\hbar\Omega/k_B T)\,S(q,\Omega)$, where the Stokes ($\Omega <
0$) and anti-Stokes ($\Omega > 0$) contributions in the spectrum are almost
equal, i.e., for small energy transfers like in quasielastic scattering or at high
temperatures. In magnetic systems, however, the condition of detailed balance
is violated because it requires the system to possess time-reversal
invariance. This is not the case in the presence of a magnetic
field \cite{SRD*09}: The equation of motion $\partial\mm/\partial t =
-\gamma \mu_0 \mm\times\mH_{eff}$ of the magnetization
$\mm$ in the effective field $\mH_{eff}$ enforces only a right-handed
precession. The time-reverted state of a left-handed precession is not
supported. Therefore one can expect a significant asymmetry in the
dynamical structure factor of magnetic systems, leading to a non-zero value for the
integral in Eq.\,(\ref{final_intens}). This asymmetry is
further enhanced with decreasing temperature. Moreover, detailed balance is
significantly violated for systems that
are strongly driven out of thermal equilibrium as it is the case, e.g., for spinwaves
that are excited by a microwave field. Since this scattering geometry with
a vertical analyser enables polarization rejection ratios up to
$10^{-10}$ in a multiple reflection geometry \cite{TAS*95,AST00,MUH*11}, a very strong
suppression of the non-resonant and non-orthogonal scattering can be achieved so that even
weak signals can be detected with good signal-to-noise ratio.  This
technique appears to be very attractive for studies at the L-edges of
the rare earths that are constituents of materials with complex and unconventional
magnetic properties and, due to their crystalline structure, should
exhibit a substantial XMLD. For these energies in the range of 6 - 9 keV one
finds Bragg reflections of Si or Ge with a Bragg angle 
close to 45$^{\circ}$ to ensure sufficiently high polarization
rejection (see supplementary material \cite{supplement}).\\
A case that will be frequently encountered in experiments is a
spinwave with a wavevector $\mq$ that lies on the dispersion surface
of the excitation with frequency  $\Omega_q$, and Lorentzian lineshape with half-width $\Gamma(q)$. We
assume that the Stokes/anti-Stokes asymmetry for that excitation can
be expressed as $S(q,\Omega) - S(q,-\Omega) =
I_0\,(\Gamma/2)^2/[(\Omega - \Omega_q)^2 + (\Gamma/2)^2])$.
Inserting this into Eq.\,(\ref{final_intens}), we obtain
\begin{equation}
I(q,t) = I_B\,S(q) + I_0\,e^{-\Gamma(q) t/\hbar}\,\sin 2\Omega_q t 
\end{equation}
The exponential results from the fanning out of the photon polarization vectors
with increasing travel distance from the sample due to the distribution of spinwave frequencies, as illustrated in Fig.\,\ref{polspiral1}a.
Experimentally, the time $t$ is translated into the travel distance
$L$ of the photons after the scattering process, i.e., $t = L/c$, so
that $I(q,t) = I(q,L/c) =: I(q,L)$ can be measured via a position
sensitive analyzer behind the sample, consisting of a strongly
asymmetrically cut crystal with a Bragg angle equal or close to
45$^{\circ}$ (the Brewster angle for hard x-rays), as illustrated in
Fig.\,\ref{polspiral1}b (see also supplementary material
\cite{supplement}).  Thus, one period $T$ of the spinwave is
mapped to a distance of $L  = \pi c/\Omega$. Assuming that the spatial
point spread function introduced by analyzer and detector has a width of $L_{min} \approx
100\,\mu$m, and that about 10 sampling points are required to resolve one modulation
period, one finds that spinwaves up to a frequency of $f_{max} =
\Omega_{max}/2\pi = c/(20 L_{min}) = 150$\,GHz can be detected. This
covers a wide range of magnetic dynamics that can be excited, e.g.,
via microwave fields.\\
It should be noted that the formalism laid out in this paper is based
on a classical description. This approach is valid either when a large number
of magnons is excited in the system, e.g., via pumping by an external
stimulus or when $k_B T \ll J S^2$ (with $J$ being the exchange
interaction constant) but $k_B T \gg \hbar\Omega$. Since $\hbar\Omega
\sim J\,S$, this can be valid only when $S \gg 1$, i.e., for large
spins. This applies for many of the rare earth (RE) elements that exhibit
large magnetic moments close to that of a free ion. At the L-edges of
RE compounds and transition metal oxides like cuprates and maganites
one often finds a relatively large XMLD contribution where this
spectroscopy relies on.  Another very interesting realization of the classical limit
are collective spinwave modes in nanoparticles in which at
sufficiently low temperatures only the $q = 0$ mode is populated where
all spins precess in unison \cite{HLL93}. These modes (that can also
be excited by microwave fields \cite{GFJ*06}) lead to peculiar
magnetic properties of antiferromagnetic nanoparticles \cite{FM04}.\\
The scheme proposed here offers the
possibility to record $I(q,t)$ for a given $q$ in a single shot at x-ray free-electron
lasers in combination with pump-probe schemes with unprecedented
energy resolution. This allows to reveal magnetic microstates as they are populated, for example,
during  magnetic switching and reversal processes. If phase locked to
a periodic excitation process, similar studies can be done in a stroboscopic fashion
already at conventional synchrotron radiation sources. The combination
with efficient micro- and nanofocusing of high-brilliance x-rays
allows to uniquely access magnetic dynamics in low-dimensional systems as they are
relevant in the field of spintronics and magnonics.
\begin{acknowledgments}
I acknowledge fruitful discussions with
Guido Meier, Lars Bocklage and Liudmila Dzemiantsova.
\end{acknowledgments}
\end{document}